\documentclass[12pt,twocolumn]{iopart}

\usepackage{graphicx}
\usepackage{bm}
\input colordvi

\begin{document}

\title[FFLO phase in the presence of pair hopping interaction]
{Fulde--Ferrell--Larkin--Ovchinnikov phase in the presence
of pair hopping interaction}

\author{Andrzej Ptok$^1$, Maciej M Ma{\'s}ka$^{1,2}$ and Marcin Mierzejewski$^1$}
\address{$^1$Department of Theoretical Physics, Institute of Physics, University of
Silesia, 40--007 Katowice, Poland}
\address{$^2$Department of Physics, Georgetown University, 37th and O Sts.
NW, Washington,~D.C.~20057, USA}

\date{\today} 
\begin{abstract}
The recent experimental support for the presence of
the Fulde--Ferrell--Larkin--Ovchinnikov (FFLO) phase
in the CeCoIn$_5$ directed
the attention towards the mechanisms responsible for this type
of superconductivity. We investigate the FFLO state 
in a model where on--site/inter--site pairing coexists with repulsive pair hopping interaction. 
The latter interaction is interesting in that it
leads to pairing with nonzero momentum of the Cooper 
pairs even in the absence of the external magnetic field (the so--called $\eta$--pairing).  
It turns out that depending on the strength of
the pair hopping interaction the magnetic field can induce one of two types of
the FFLO phase with different spatial modulations of the order parameter.
It is argued that the properties of the FFLO phase may give information about
the magnitude of the pair hopping interaction. We also show that $\eta$--pairing and
$d$--wave superconductivity may coexist in the FFLO state. It holds true also
for superconductors which in the absence of magnetic field are of pure $d$--wave type.
\end{abstract}

\submitto{\JPCM}
\maketitle

\section{Introduction}

An unconventional superconducting state with a non--zero
total momentum of the Cooper pairs was predicted by
Fulde and Ferrell \cite{FF} as well as by Larkin and Ovchinnikov \cite{LO}
in the middle of 60's. Under particular conditions, 
this phase should occur at low temperatures and 
in strong magnetic fields.
Due to sever requirements for the 
formation of the FFLO state, this type of superconductivity
has experimentally been observed only recently.
In the FFLO state the superconducting order parameter (OP) oscillates
in the real space. This property, to some extend, resembles the unconventional
superconductivity in strongly correlated systems \cite{matsuda}, where   
the OP changes sign in the momentum
space. 
The FFLO state has recently been analyzed 
in the context of heavy fermion systems \cite{hf1,hf2,hf3,hf4,hf5,hf6,hf7,hf8,
hf9,hf10,hf11,hf12,hf13,hf14,hf15},
organic superconductors \cite{org_sup1,org_sup2}, ultracold atoms \cite{uca1,uca2} and a dense
nuclear matter \cite{dnm1,dnm2,dnm3}.
Although, there is no direct experimental evidence for the spatial variation of the OP,
suggestions for the future experiments have been developed in Refs. \cite{hf6,se1,se3}. 

The orbital (diamagnetic) pair breaking is a crucial mechanism that 
limits realization of the FFLO state.
In the vast majority of superconducting materials it
is a dominating pair breaking mechanism that destroys
superconductivity for magnetic fields much weaker
than the Clogston-Chandrasekhar limit ($H^{CC}$) \cite{CC1,CC2}.
It holds true also for models appropriate to describe the short
coherence length superconductors \cite{my1,my2,my3}.
The significance of the diamagnetic pair breaking
is usually described in terms of the
Maki parameter $\alpha=\sqrt{2}H^{\mathrm orb}_{c2}/H^{CC}$, 
where $H^{\mathrm orb}_{c2}$ is the upper critical field calculated
without the Zeeman splitting.  There exist two general possibilities to reduce
the destructive role of the orbital pair breaking.
In the layered superconductors, formation of Landau orbits should be suppressed
for magnetic fields applied parallel to the layers.
This may explain possible observations of the FFLO state in some 
organic superconductors \cite{org_sup1,org_sup2}. The role of the orbital pair breaking should also 
be limited in systems with narrow energy bands, like heavy fermion
systems. The experimental evidence for the FFLO superconductivity
in these systems seems to the strongest \cite{hf1,hf2,hf3,hf4,hf5,hf6,hf7,hf8,hf9,hf10,hf11,hf12,hf13}.

In the context of recent investigations of the FFLO state, it is important to
search for other mechanisms that stabilize superconductivity   
against the orbital pair breaking.
Recently it has been found that superconductivity originating 
from repulsive pair hopping interaction is unique in that it is robust
against this pair breaking mechanism \cite{my}.
The origin of this interaction may vary in different systems and therefore we do not
specify a particular superconductor for which the following qualitative analysis can directly be applied.
The repulsive pair hopping interaction can be derived from a general microscopic
tight--binding Hamiltonian \cite{hubbard}, but 
in this case the magnitude of the interaction is very small. 
However, since this interaction leads to superconductivity that is almost unaffected 
by the orbital effects, it may become more important close to the upper critical field,
i.e., in the regime where the FFLO phase is expected to occur.
We may also consider other sources of the pair hopping  which give rise to much larger 
magnitudes of this interaction. 
For example, such term may be included in the effective Hamiltonian describing Fermi gas 
in optical lattice in the strong interaction regime \cite{ph-opt-lat1, ph-opt-lat2}. 
It also arises in a natural way in multiorbital models \cite{oles}, 
though then the pairs hop between different orbitals on the same site. 
Nevertheless, we expect that some of our conclusions can still be valid. The role 
of the pair hopping interaction in a multiorbital models is of particular interest 
because of its presence in the recently discovered iron--pnictides \cite{pnictides}.
Additionally, if the on--site repulsion exceeds the gap
between the lowest and the next--lowest bands in optical lattice, then the interband 
pair hopping seems to
important. Very recent Quantum Monte Carlo calculations for TMTSF-salt \cite{TMTSF} 
also suggest a significant role of the pair hopping processes in this system, which, 
on the other hand, probably exhibits the FFLO phase at high field 
\cite{org_sup2,TMTSF-FFLO3}.

In the absence of magnetic field, the pair hopping interaction is responsible for the
$\eta$--type pairing where
the total momentum  of the paired electrons is ${\bm Q}=(\pi,\pi)$
and the phase of superconducting order parameter alters
from one site to the neighboring one \cite{pens,rob-bul,czart1,czart2,japaridze}.
It has been shown that flux quantization and Meissner effect
appear in this state \cite{yang1,yang2}.
Therefore, even in the absence of external magnetic field,
the repulsive pair hopping interaction favors
pairing with non--zero momentum of Cooper pairs. 
Although, the pair hopping may not be the dominating pairing mechanism,
its presence may affect the FFLO phase. In the following, the superconducting
state with zero total momentum of the Cooper pair will be referred to as 
the BCS superconductivity. In that sense, both FFLO and $\eta$--pairing
will be considered as a non--BCS state.    

In the present paper, we analyze the role of the repulsive pair hopping
interaction for the FFLO state. 
The above mentioned systems, where the pair hopping interaction may play a 
significant role, exhibit both inter--site pairing
(CeCoIn$_5$, TMTSF) as well as on--site pairing (optical lattice, the iron--pnictides). 
Therefore, we consider a model where this interaction
coexists with on--site or inter--site pairing potential.
However, we do not refer to any particular system. As the microscopic
mechanism of superconductivity in most of the unconventional 
superconductors is still under debate, we do not discuss the origin of the pairing 
potentials.

For the system with on--site pairing,  magnetic field
reduces the total momentum of electrons forming Cooper pairs in the 
$\eta$--pairing state. As a consequence, the amplitude of the superconducting 
order parameter becomes a site dependent quantity. 
Recent theoretical investigations of the FFLO phase in the attractive Hubbard model 
have been motivated mostly by the increasing interest in the ultracold Fermi gases \cite{rizzi,koponen}.
These approaches may also be applicable to compounds other than the strongly correlated heavy fermion systems
\cite{chiao1,chiao2}.
Contrary to this, experimental results obtained for the CeCoIn$_5$ indicate on the anisotropic $d$--wave pairing. 
Therefore in Sec. IV. we study the case of 
inter--site attractive interaction that is responsible for the $d$--wave superconductivity.
We show for the case of inter--site attraction that 
$d$--wave and $\eta$--pairing orders may coexist in the FFLO state. Such a coexistence is possible also for
systems which in the absence of magnetic field are in the pure $d$--wave state.
For on--site and inter--site pairings, the potential of the pair hopping
interaction $J$ is assumed to be positive.

\section{ On--site pairing}  

 We start our analysis with a model with on--site pairing interaction
described by the following tight--binding Hamiltonian:

\begin{eqnarray}
H &=& -t \sum_{\langle i,j \rangle , \sigma } c_{i \sigma}^{\dagger} c_{j \sigma} 
+ U \sum_{i} c_{i \uparrow}^{\dagger} c_{i \uparrow} c_{i \downarrow}^{\dagger} 
c_{i \downarrow} \nonumber \\ 
&+& J \sum_{\langle i,j \rangle} c_{i \uparrow}^{\dagger} c_{i \downarrow}^{\dagger} 
c_{j \downarrow} c_{j \uparrow} - \sum_{i , \sigma} \left[s(\sigma) h  + \mu \right] 
c_{i \sigma}^{\dagger} c_{i \sigma},
\label{h1}
\end{eqnarray}
where $t$ is the nearest neighbor hopping integral,  $J$ is the pair hopping interaction, $\mu$ is the  chemical potential,
$s(\uparrow)=1$, and $s(\downarrow)=-1$.
The Zeeman coupling is determined by $h=g \mu_B {\cal H} / 2$, 
where $g$ is the gyromagnetic ratio,
$\mu_B$ is the Bohr magneton and $\cal H$ is the external magnetic field.
Hamiltonian (\ref{h1}) does not include  the diamagnetic pair breaking. 
We refer to Refs. \cite{diama1,diama2,diama3,diama4,diama5,diama6,diama7,diama8,diama9,diama10} for the discussion 
of the influence of this mechanism 
on the FFLO state.
Here, we focus on the role of the pair hopping interaction as well as on the properties of
the $\eta$--pairing, that is robust against the orbital pair breaking.
Nevertheless, the role of the orbital pair breaking  will be briefly discussed.
In this section we assume the simplest
form of the {\em effective} on--site pairing interaction ($U<0$), that is responsible for
the $s$--wave superconductivity.

We apply the mean field approximation and assume the order parameter in a general
form:
\begin{equation}
\Delta(\bm R_j) \equiv \langle c_{j\downarrow} c_{j \uparrow}  \rangle= \sum_{m=1}^{M} \Delta_{m} \exp ( i {\bm Q}_{m} \cdot {\bm R}_{j} ). \label{parporz}
\end{equation}
Then, the Hamiltonian in the momentum  space takes the form
\begin{eqnarray}
H_{MF} &=& \sum_{k \sigma} \tilde{\varepsilon}_{{\bm k} \sigma} c_{{\bm k} \sigma}^{\dagger} c_{{\bm k} \sigma} 
\nonumber \\
&+& \sum_{m=1}^{M} U_{\mathrm{eff}} ( {\bm Q}_{m} ) \sum_{\bm k}\left( \Delta_{m}^* c_{-{\bm k}+{\bm Q}_{m} \downarrow} c_{{\bm k} \uparrow} + \mathrm{ H.c.} \right) \nonumber \\
&-& N \sum_{m=1}^M  U_{\mathrm{eff}} ( {\bm Q}_{m} ) |\Delta_m|^2 
\end{eqnarray}
where
\begin{equation}
\tilde{\varepsilon}_{{\bm k} \sigma} = \varepsilon_{\bm{k}} - \mu - s(\sigma) h.
\end{equation}
For arbitrary lattice geometry the dispersion relation is given by 
 $\varepsilon_{\bm{k}}=-t/N\sum'_{j}\exp \left(i \bm{R}_j \cdot \bm{k} \right)$, where the prime means summation over the nearest neighbor sites. We have also introduced
an effective pairing potential
\begin{equation}
U_{\mathrm{eff}} ( {\bm Q} ) = U - \frac{J \varepsilon_{\bm Q}}{t}.
\end{equation}

For a general form of the order parameter [Eq. (\ref{parporz})] 
diagonalization of the mean field  Hamiltonian  usually cannot be reduced to
an eigenproblem of a finite Hermitian matrix. Therefore,
we restrict further discussion to two simplest cases. The first one
was originally proposed by Fulde and Ferrel (FF),
whereas the second by 
Larkin and Ovchinnikov (LO). 
In the former one, it is assumed that $M=1$, so the absolute value
of $\Delta(\bm R_j)$ is constant, but the phase changes from site to site.
In the latter case $M=2$, $\Delta_1=\Delta_2$ and $\bm Q_1=-\bm Q_2$.
Then, one gets $\Delta(\bm R_j)=2 \Delta_0 \cos(\bm Q \cdot {\bm R}_j)$, where 
we use $\Delta_0 \equiv \Delta_1$ and $\bm Q \equiv \bm Q_1$.
However, one should keep in mind that at low temperature and high magnetic field
FFLO phases with $M>2$ may be thermodynamically more stable \cite{shi,mora}.

For the FF phase, the mean field Hamiltonian can be diagonalized by means of the
Bogoliubov transformation. Straightforward calculations lead to the following
form of the grand canonical potential $\Omega= -kT \ln {\rm Tr} \exp(-\beta H) $
\begin{eqnarray}
\Omega &=& - k T \sum_{\alpha = \pm } \sum_{\bm k} \ln \left[ 1 + \exp ( - \beta 
E_{\bm k,\alpha} ) \right]  \nonumber  \\
&+& \sum_{\bm k} \tilde{\varepsilon}_{-{\bm k}+{\bm Q} \downarrow} - NU_{\mathrm{eff}}({\bm Q}) | \Delta_{0} |^{2},
\end{eqnarray}
where
\begin{eqnarray}
E_{\bm k,\pm} &=& \frac{1}{2} \left[ \: \tilde{\varepsilon}_{\bm k\uparrow} 
- \tilde{\varepsilon}_{-\bm k+{\bm Q}\downarrow}  \right. \nonumber \\
& \pm & \left. \sqrt{ \left( \tilde{\varepsilon}_{\bm k\uparrow} 
+ \tilde{\varepsilon}_{-\bm k+{\bm Q}\downarrow} \right)^{2} + 4U_{\mathrm{eff}}({\bm Q})^{2} | \Delta_{0} |^{2} }\ \right].
\end{eqnarray}

In the case of LO superconductivity, the Hamiltonian can not be
diagonalized analytically. However, the pairing term 
links the one--particle states with momenta lying along a single
line in the Brillouin zone. Therefore, one can solve the resulting
eigenproblem numerically for relatively large clusters. Namely,
for a $L \times L $ cluster, one has to diagonalize a $2L \times 2L$
Hermitian matrix.

\subsection{Numerical results} 
\begin{figure}
\includegraphics{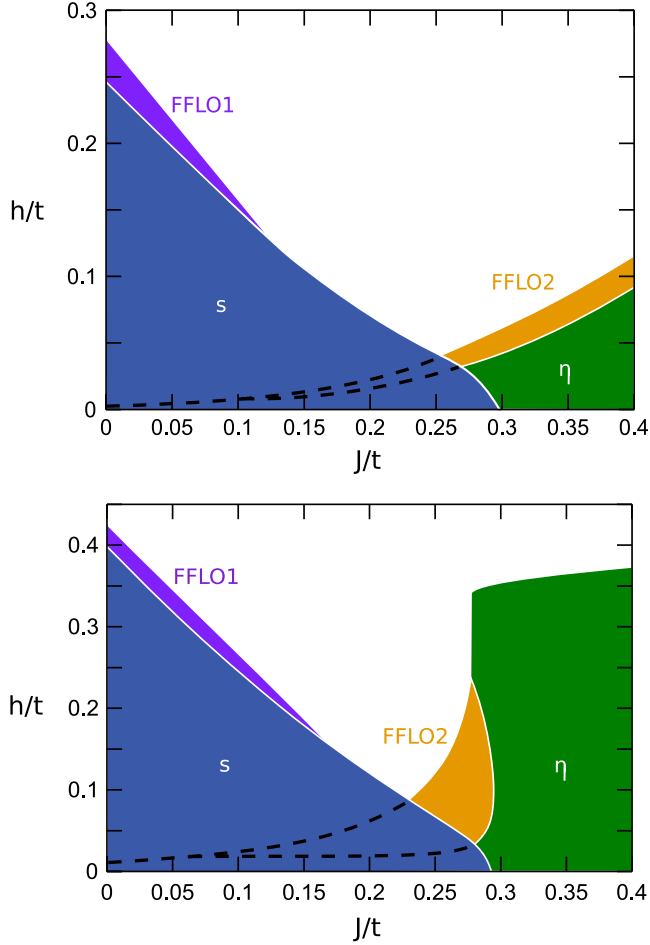}
\caption{Phase diagram showing the stable superconducting phases
for various $J$ and $h$ at $T=0$. The upper panel shows results
obtained for a square lattice with $U = -2.0t$, whereas the lower
one shows results for a triangular lattice with $U = -2.5t$. 
The 
dashed lines are explained in the text.}
\label{fig1}
\end{figure}

\begin{figure}
\includegraphics{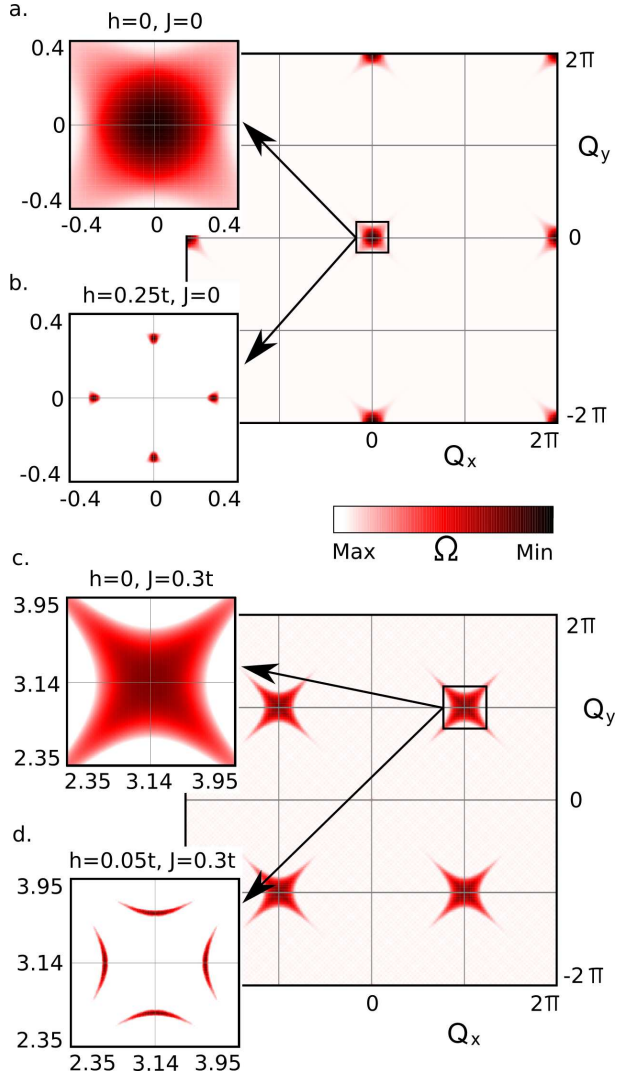}
\caption{Grand canonical potential minimized with respect to $|\Delta_0|$ for
given $\bm Q=(Q_x,Q_y)$. We have used the same parameters $\mu$ and $U$ as
in Fig. \ref{fig1}, whereas the values of $h$ and $J$ 
are shown explicitly above the small panels. 
Note different ranges of $Q_x$ and $Q_y$ in various panels. These results
have been obtained for a square lattice.}\
\label{fig2}
\end{figure}
\begin{figure}
\includegraphics{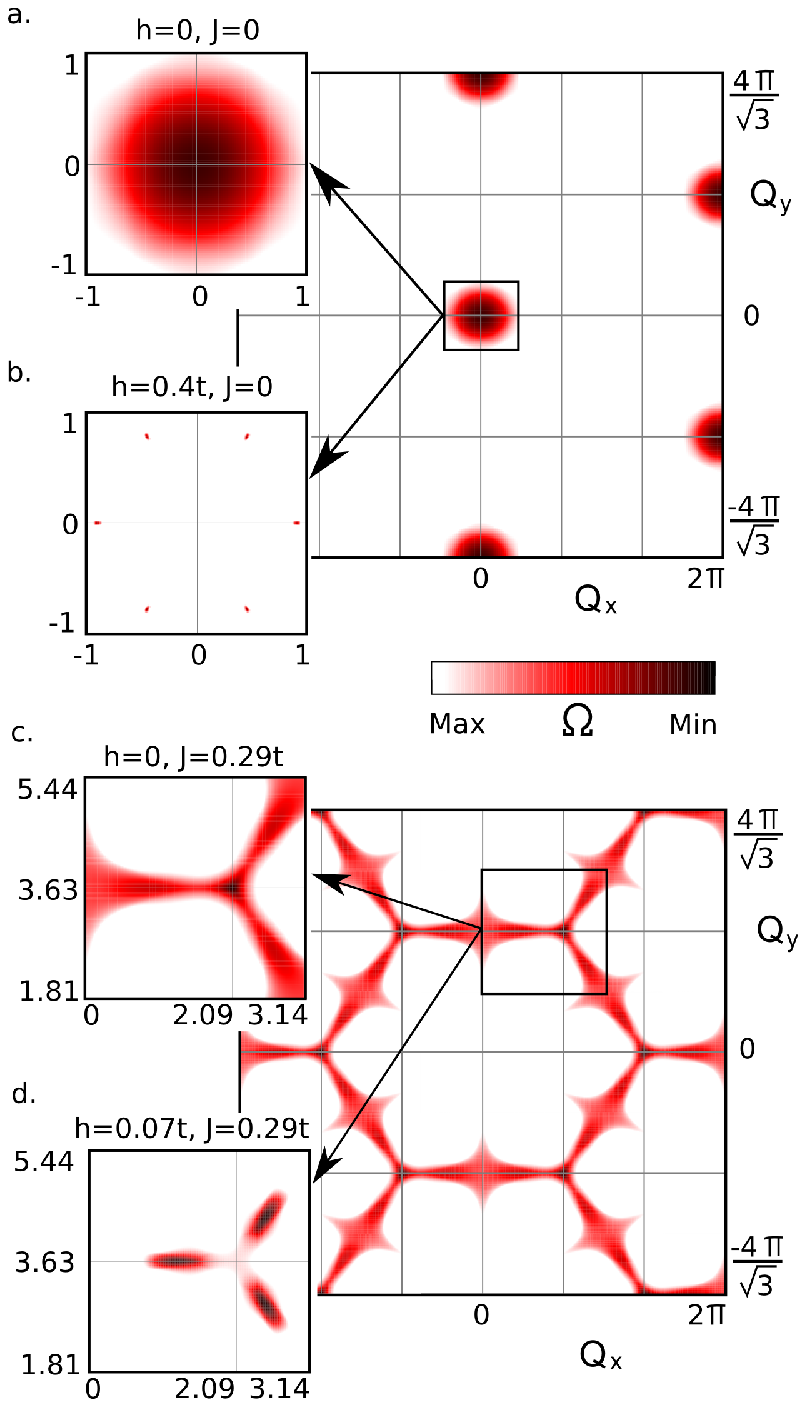}
\caption{The same as in Fig. \ref{fig2} but for a triangular lattice.}
\label{fig3}
\end{figure}

We start our discussion with the simplest case $M=1$ (FF state), that allows one to
estimate the boundaries of the non-BCS superconducting phases. However, 
the presence of the LO superconductivity will be discussed as well.
The thermodynamically stable phase has been determined through
minimization of the grand canonical potential with respect to
the superconducting order parameter $|\Delta_0|$ and $\bm Q$. The 
calculations have been carried out for a square lattice with
$\mu=0$ as well as for a triangular lattice with $\mu=2t$. 
These chemical potentials correspond to maxima in
the density of states and, therefore, to the highest 
superconducting transition temperatures. 
A comparison of results obtained for both
the cases allows one to check the role of the lattice geometry.
      
\begin{figure}
\includegraphics{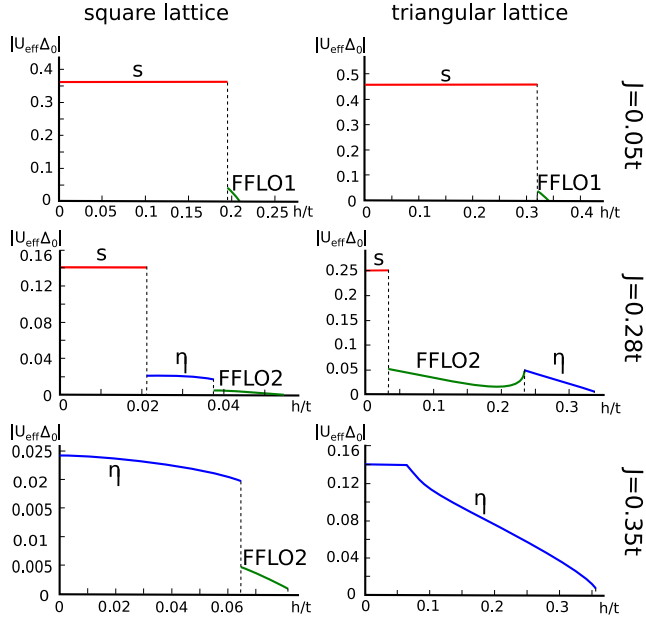}
\caption{$|U_\mathrm{eff}({\bm Q}) \Delta_0|$ as a function of magnetic field. The parameters as well as
the lattice geometry are explicitly shown in the figure. }
\label{fig4}
\end{figure}

First, we focus on the influence of the pair hopping interaction
on the properties of the FFLO phase. Fig. \ref{fig1} shows how the ground
state of the system depends on $J$ and $h$. In Figs. \ref{fig2} (square lattice) 
and \ref{fig3} (triangular lattice) 
we present the values of $\bm Q$ that minimize $\Omega$
for some particular values of  $J$ and $h$.  
In the absence of  the magnetic field there are two stable superconducting phases
for both the lattice geometries. For small $J$ there exists
an isotropic BCS phase, that will be referred 
to as  $s$--wave superconductivity (see Figs. \ref{fig2}a and \ref{fig3}a). 
The $\eta$--pairing
phase occurs for larger $J$. In the case of a bipartite square lattice,
$\eta$--pairing  corresponds to the total momentum of Cooper pairs 
$\bm Q=(\pi,\pi)$, where the phase of superconducting order
parameter changes from one lattice site to the neighboring one 
(see Fig. \ref{fig2}c).
Stability of this phase obviously follows from the fact, that 
the pair hopping interaction involves sites, which belong to
different sublattices. Then,
the oscillating $\Delta_i$ minimizes the energy of the system,
for the physically relevant repulsive interaction.
In this context, the problem of $\eta$--pairing on a non--bipartite triangular
lattice is interesting even in the absence of magnetic field \cite{aptok1}.
We have found that also for this geometry,
the repulsive pair hopping interaction may lead to thermodynamically
stable phase with $\bm Q \neq 0$.  The ground state energy is minimal when $\bm Q$ 
represents one of the corners of the hexagonal first Brillouin zone 
(see Fig. \ref{fig3}c).
It is easy to check that for such a value of $\bm Q$, the phase of $\Delta_i$ takes on
three different values, namely,  $\Delta_{i}=\Delta_{0}$, $\Delta_{i}=\Delta_{0} \exp ( i \frac{2}{3} \pi )$ or
$\Delta_{i}=\Delta_{0} \exp ( - i \frac{2}{3} \pi )$ depending on $i$.
In the presence of sufficiently strong magnetic field there exist four different
superconducting phases. Apart from the discussed above $s$--wave and
$\eta$--pairing states there are  other two phases, that will be
referred to as FFLO1 and FFLO2. Investigation of the total momenta
of Cooper pairs (see panels ``b'' and ``d'' in Figs. \ref{fig2} and \ref{fig3}), 
allows one to link FFLO1 and FFLO2 to $s$--wave and $\eta$--pairing, respect what, in turn, is a hallmarkively.
Namely, $\bm Q$ obtained for the FFLO1 is relatively close to the origin of
the Brillouin zone, whereas in the FFLO2 phase $\bm Q$ remains on the edges
of the zone. One can see, that FFLO2 phase occurs for lower magnetic fields
than the FFLO1. FFLO1 evolves from the
s-wave superconductivity when sufficiently strong magnetic field
is applied. It holds true for both the lattice geometries. 
On a square lattice, FFLO2 evolves from the $\eta$--pairing state
under the same conditions.
However, on a triangular lattice this phase
is stable only for moderate values of the pair hopping interactions as well
as for moderate magnetic fields. It is surprising that increasing of
magnetic field may cause two phase transitions, the first one from $s$--wave
to FFLO2 is discontinuous and the second from FFLO2 to $\eta$--pairing is a continuous
transition (see Fig. \ref{fig4}). We have also found that
the transitions from the s-wave phase to the FFLO1 state
are discontinuous, whereas the transitions from the FFLO1 and FFLO2 phases
to the normal state are continuous. Finally,  for a square lattice the transition
from the $\eta$--pairing state to the FFLO2 phase is discontinuous as well.
The field dependence of the superconducting order parameter in various
phases can inferred from  Fig. \ref{fig4}.      

In the absence of the pair hopping interactions, it is well known that 
the LO phase has lower ground state energy than the FF one \cite{LO}.
In our case,  this result directly applies to the FFLO1 phase. We have found that 
in the case of FFLO2, state with $M=2$ has energy lower than that with $M=1$.
Therefore, also in this case $|\Delta_i|$ is spatially inhomogeneous.
This result has been obtained on the basis of numerical diagonalization 
of $200 \times 200$ clusters  with periodic boundary conditions.
Consequently, there exists a simple criterion to distinguish between FFLO1
and FFLO2 phases. In both the cases $\Delta_i \sim \cos \bm Q \cdot \bm R_i $
but $|\bm Q| \ll 1$ for FFLO1, whereas $|\bm Q|  > \pi$  for FFLO2.
Therefore, the periods of spatial modulations of the order parameters relevant
to FFLO1 and FFLO2 are very different. In the latter case it is of the order
of the lattice constant.  
It is instructive to examine in more detail the spatial modulation of $\Delta_i$ in the FFLO2 phase on a square lattice.
As the total momentum of Cooper pairs ${\bf Q}$ is close to ${\bm \Pi}=(\pi,\pi)$, one can introduce 
 ${\bf Q'}={\bm \Pi}-{\bm Q}$ and note that  $|\bm Q'| \ll 1$. Then, $\Delta_i \sim \cos [\bm (\bm \Pi - \bm Q') \cdot \bm R_i]=
\cos (\bm \Pi \cdot \bm R_i)\cos (\bm Q' \cdot \bm R_i) $. The spatial profile of $\Delta_i$ is determined by two oscillating
functions. Due to the first one the phase of the superconducting order parameter alters from one site to the neighboring one. It means that the FFLO2 phase retains the
basic properties of the $\eta$--pairing superconductivity. The second
factor is responsible for a slow variation of the magnitude of superconducting order parameter $|\Delta_i|$ what is a hallmark of the
LO--type of superconductivity.
On the basis of our analysis one can not exclude
that the FFLO2 phase with $M > 2$ is more stable.  
Therefore, the actual boundaries of the FFLO2 state may cover a slightly wider range of  
magnetic fields, than presented in Fig. \ref{fig1}. 
    
\begin{figure}
\includegraphics{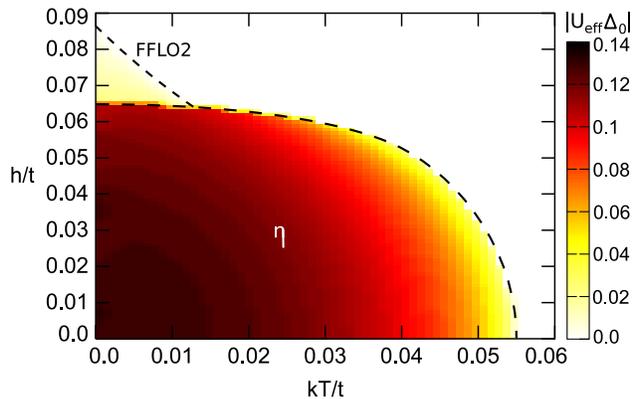}
\caption{$|\Delta_0|$ as a function of magnetic field and temperature for
a square lattice with $J=0.35t$.}
\label{fig6}
\end{figure}
  
Up to this point we have analyzed a two--dimensional system, where the influence 
of the orbital pair breaking can be neglected provided the applied magnetic field is
parallel to the plane. However, if the magnetic field has a nonzero component perpendicular
to the plane, as well as in the case of three dimensional systems the destructive
role of the orbital effects has to be taken into account. This problem has been analyzed, 
e.g., in Refs. 
\cite{my1,my2,my3,my,diama1,diama2,diama3,diama4,diama5,diama6,diama7,diama8,diama9,diama10}. 
On the one hand, it is the dominating pair--breaking mechanism for $s$--wave and FFLO1 superconductivity \cite{my1,my2,my3,diama1,diama2,diama3,diama4,diama5,diama6,diama7,diama8,diama9,diama10}. On the other 
hand, it is very ineffective in destroying $\eta$--type superconductivity \cite{my}.
Therefore, we assume that FFLO2 will also be robust against the 
orbital pair breaking. This property may lead to significant modifications of the phase
diagram presented in Fig. \ref{fig1}. Namely, the dashed lines show the boundaries
of the $\eta$ and FFLO2 types of  superconductivity obtained under the assumption
that $s$--wave and FFLO1 phase are destroyed by orbital effects.
Note, that even weak pair hopping interaction should lead to the onset of
$\eta$ and FFLO2 phases for fields sufficiently strong to destroy the
conventional superconductivity. However, this conjectural result should 
be confirmed by calculations for
the FFLO2 phase with the diamagnetic 
pair breaking explicitly taken into account.
  
Finally, we discuss the standard $(kT,h)$ phase diagram for 
a square lattice with $J=0.35t$.
It is the value of pair hopping interaction
for which the ground state in the absence of magnetic field is the $\eta$--type
superconductivity (see Fig. \ref{fig1}).  Fig. \ref{fig6} shows the results.  
It is interesting that despite unconventional
character of the $\eta$--pairing, the phase diagram is very similar to analogous
phase diagram for BCS--FFLO superconductors. Namely, the FFLO2 phase occurs only in
the presence of strong magnetic field and at low temperatures. The phase
transition from $\eta$ to FFLO2 phase is discontinuous, whereas the transition
from FFLO2 to the normal state is continuous.   

\section{Inter--Site pairing}


 In this section we extend our previous study by allowing for the anisotropic $d$--wave superconductivity.
For this sake, we add the nearest--neighbor attraction term to the Hamiltonian (\ref{h1})
\begin{equation}
H\rightarrow H+
V \sum_{\langle i,j \rangle, \sigma} c_{i \sigma}^{\dagger} c_{i \sigma}
c_{j, -\sigma}^{\dagger} c_{j, -\sigma},
\label{hdwave}
\end{equation}
where we assume $V<0$. For simplicity, we restrict ourselves to FF superconductivity on a square lattice.
Then, the mean--field Hamiltonian in the momentum space reads
\begin{eqnarray}
H_{MF} &=& \sum_{k \sigma} \tilde{\varepsilon}_{{\bm k} \sigma} c_{{\bm k} \sigma}^{\dagger} c_{{\bm k} \sigma} 
\nonumber \\
&+&  \sum_{\bm k}\big\{ \left[ U_{\mathrm{eff}} ( {\bm Q} ) \Delta_0^* + V d ( {\bm k} ) \Delta_d^* \right] c_{-{\bm k}+{\bm Q} \downarrow} c_{{\bm k} \uparrow}  \nonumber \\
&+&  \mathrm{ H.c.} \big\} - N U_{\mathrm{eff}} ( {\bm Q} ) |\Delta_0|^2 - 2 N V |\Delta_d|^2,
\label{mdmf}
\end{eqnarray}
where
$\Delta_d=1/(2N) \sum_{\bm k} d({\bm k}) \langle  c_{-{\bm k}+{\bm Q} \downarrow} c_{{\bm k} \uparrow} \rangle $ and
 $d ( {\bm k} )= 2 ( \cos k_x - \cos k_y)$.
 We have assumed that the expectation value 
$\langle c_{-{\bm k}+{\bm Q} \downarrow} c_{{\bm k} \uparrow} \rangle$ is non--zero only for one
particular wave vector $\bm Q$.
In clear contrast to case discussed in the preceding section, one has to introduce 
two superconducting order parameters $\Delta_0$  and $\Delta_d$
originating from the on--site pairing and the inter--site attraction, respectively.  
One can straightforwardly calculate the grand canonical potential that is of the form
\begin{eqnarray}
\Omega &=& - k T \sum_{\alpha = \pm } \sum_{\bm k} \ln \left[ 1 + \exp ( - \beta 
E_{\bm k,\alpha} ) \right] + \sum_{\bm k} \tilde{\varepsilon}_{-{\bm k}+{\bm Q} \downarrow} \nonumber  \\
&-& N U_{\mathrm{eff}}({\bm Q}) | \Delta_{0} |^{2} -2 N V |\Delta_d|^2 
\end{eqnarray}
with the  quasiparticle energies $ E_{\bm k,\pm}$
\begin{eqnarray}
E_{\bm k,\pm} &=& \frac{1}{2} \bigg\{  \tilde{\varepsilon}_{\bm k\uparrow} 
- \tilde{\varepsilon}_{-\bm k+{\bm Q}\downarrow} \\
& \pm & \sqrt{ \left( \tilde{\varepsilon}_{\bm k\uparrow} 
+ \tilde{\varepsilon}_{-\bm k+{\bm Q}\downarrow} \right)^{2} 
+ 4 | U_{\mathrm{eff}}({\bm Q}) \Delta_{0}  + V d ( {\bm k} ) 
\Delta_d |^{2}  } \bigg\} . \nonumber 
\end{eqnarray}

\subsection{Numerical results}

 The presence of two order parameters $\Delta_0$ and $\Delta_d$ makes the problem numerically much more 
complicated than for  the on--site pairing potential only. Though the coexistence of the 
$s$--wave and $d$--wave pairings in the FFLO phase may significantly enhance the upper
critical field \cite{jin}, for the sake of simplicity and to avoid too many model parameters 
we restrict our further analysis to the case $U=0$.

\begin{figure}
\includegraphics{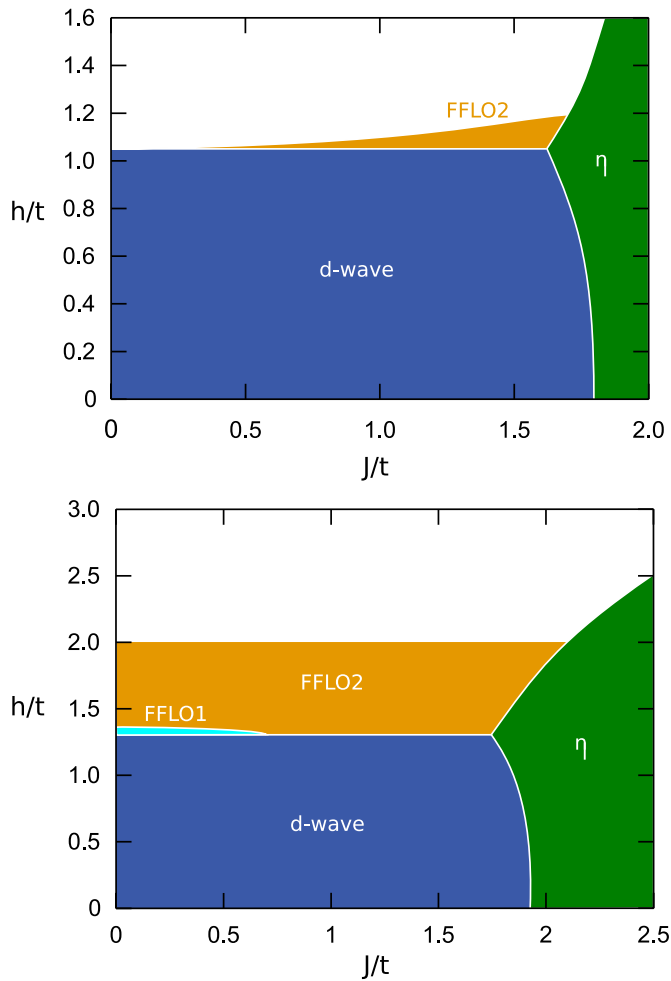}
\caption{
Phase diagram for a square lattice showing the stable superconducting phases
for various $J$ and $h$ at $T=0$. The upper panel shows results
obtained for $V = -2.0t$, whereas the lower
one shows results for $V = -2.5t$.}
\label{fig6_new}
\end{figure}

\begin{figure}
\includegraphics{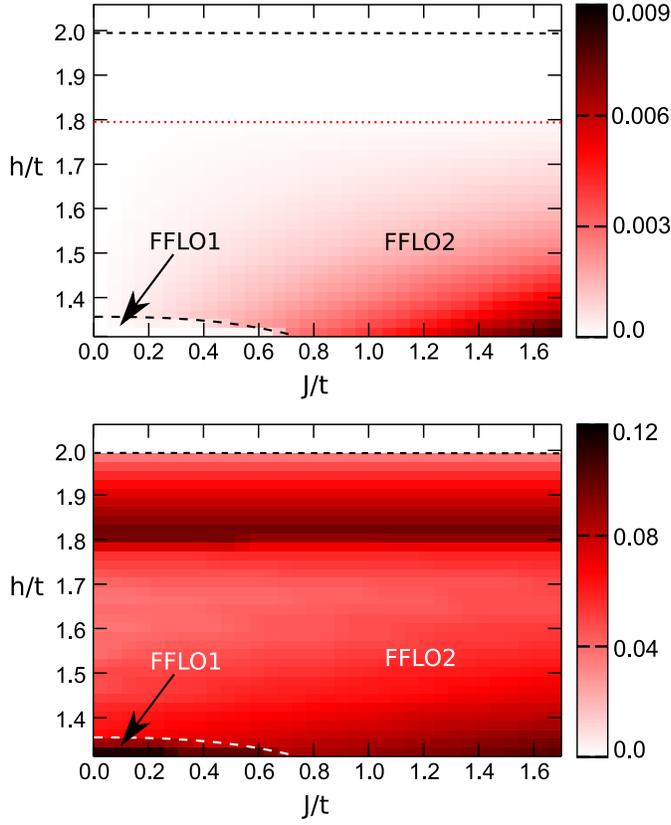}
\caption{
 $|U_\mathrm{eff}({\bm Q})\Delta_0|$  (upper panel) and $|V \Delta_d|$ (lower panel) 
for various $J$ and $h$ at $T=0$. $V=-2.5t$ has been assumed. Dashed lines
show show the boundaries of the FFLO1 and FFLO2 phases. In the upper
panel  $|U_\mathrm{eff}({\bm Q})\Delta_0|$ vanishes above the dotted line. 
}
\label{fig7_new}
\end{figure}

\begin{figure}
\includegraphics{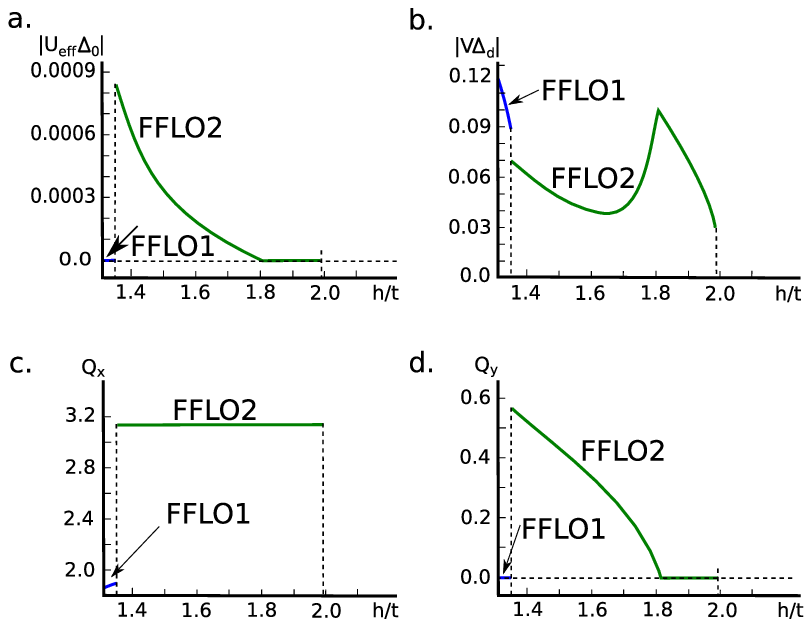}
\caption{Magnetic field dependence of  $|U_\mathrm{eff}({\bm Q})\Delta_0|$  (a),
 $|V \Delta_d|$ (b), $Q_x$ (c), $Q_y$ (d) for $T=0$, $V=-2.5t$ and  $J=0.3t$. }
\label{fig8_new}
\end{figure}

Generally, the grand canonical potential should be minimized with respect to five variables:
two components of the wave vector $\bm Q$, magnitudes of two order parameters, and the relative
phase $\phi$ between $\Delta_0$ and $\Delta_d$. If the orders do not coexist,
i.e., either  $\Delta_0$ or $\Delta_d$ vanishes, $\Omega$ is independent of  $\phi$.
However, in order to analyze whether the coexistence is possible, we have taken $\phi \in \{0,\pi,\pm \pi/2 \}$ and
minimized $\Omega$ with respect to remaining variational parameters. The resulting phase diagram
is presented in Fig. \ref{fig6_new}. 
Of course, the exact boundaries could be a bit different from those presented in this figure if
one allows for an arbitrary value $\phi$.

Sufficiently strong magnetic field drives the system into
the FF state. Depending on the values of $\bm Q$, 
one can distinguish between two phases marked in Fig. \ref{fig6_new}
as FFLO1 and FFLO2. In the former case  $\Delta_d \neq 0$, ${\bm Q}=(0,Q_y)$  and $Q_y \ll \pi $.
Then, $U_\mathrm{eff}({\bm Q})$ is positive and $\Delta_0=0$. However, in the FFLO2 state
${\bm Q}=(\pi,Q_y)$,  $Q_y \ll \pi $. We have found that in the FFLO2 phase {\em both the order parameters
may simultaneously be non--zero}. It strongly contrasts with the  results obtained in the absence of 
magnetic field, when
the system is either in the pure  $d$--wave superconducting state or in the pure $\eta$--pairing state.
Note that for moderate (presumably realistic) values of $J$ 
and in the absence of magnetic the ground state is of purely $d$--wave type. 

In order to study the coexisting orders in more detail, we have calculated $|U_\mathrm{eff}({\bm Q})\Delta_0|$ 
(see the upper panel in Fig. \ref{fig7_new}) and $|V \Delta_d|$ (see the lower panel in Fig. \ref{fig7_new})
in the FFLO2 phase. In Fig. \ref{fig8_new} we present these data for $J=0.3t$ together with
the field dependence of the wave vector $\bm Q$. Although the dominating contribution
to the superconducting gap comes from the $d$--wave pairing, 
$\Delta_0$ is non--negligible in the FFLO2 phase.  
For sufficiently strong inter--site pairing the boundaries of the FFLO2 phase are almost independent of $J$,
what can be inferred from the lower panel in Fig. \ref{fig6_new}. This result can be explained in the 
following way:  increase of the magnetic field shifts the wave vector $\bm Q$ and, in this way, 
modifies $U_\mathrm{eff}({\bm Q})$. This potential may eventually vanish causing $U_\mathrm{eff}({\bm Q})\Delta_0=0$
(see panel ``a'' in Fig. \ref{fig8_new}).
Then, the upper critical field is determined  only by the inter--site pairing. 
This effect may also be responsible for a non--monotonic field dependence of $\Delta_d$.
Neither this non-monotonicity nor $J$--independent upper critical field  
occur for weaker inter--site attraction, shown in the upper panel of Fig. \ref{fig6_new}.
Here, both the order parameters are non--zero within the entire FFLO2 phase.

\section{Concluding remarks}

Our aim was to investigate the role of the pair hopping interaction
for the FFLO superconductivity. 
Probably, this interaction is not a dominating pairing mechanism. However, 
as we have argued in the introduction, in some
superconducting systems it may become important in the high--field regime whether the FFLO phase is expected.
Motivated by the pairing symmetry in these systems we 
have separately studied two models
where the pair hopping interaction coexists with on--site and
inter--site attractions.
In the former case, the pair hopping interaction 
lowers the magnetic field corresponding to the onset of the FFLO state. 
In the presence of the inter--site pairing, sufficiently strong magnetic field allows for 
a coexistence of $d$--wave and $\eta$--pairing states even though
such a coexistence does not occur in the absence of magnetic field. 
It is instructive to compare this result with the recent
experimental and theoretical data concerning the coexistence
of superconductivity and spin--density wave in CeCoIn$_5$ \cite{sdwfflo1,sdwfflo2,sdwfflo3}.
One may formulate a general conjecture, that field--induced breaking of the
translational invariance of the superconducting phase gives way to other
competing orders.

For sufficiently strong pair hopping interaction 
one may expect $\eta$--pairing state, that is robust against the diamagnetic pair
breaking.  
According to our best knowledge this phase has not been identified
in any known superconducting system. However, upon applications
of external magnetic field such a system should exhibit
FFLO state. In contradistinction to the BCS type of pairing, this phase
should occur independently of the band--width and the orientation of magnetic field.
Investigating spatial modulation of the superconducting order parameter,
one can distinguish, whether the FFLO phase originates from BCS    
or $\eta$--pairing. In the first case the period of modulation is much
larger than the lattice constant. In the latter case it is of the 
order of the lattice constant and the phase of the order parameter retains its oscillating character
typical for the $\eta$--pairing superconductivity. 
 
\ack

The authors are grateful to Stanis{\l}aw Robaszkiewicz for a fruitful discussion.
This work has been supported by the Polish Ministry of Education 
and Science under Grants No. 1~P03B~071~30 and No. N~202~131~323786.
M.M.M. also acknowledges a support under ARO Award W911NF0710576
with funds from the DARPA OLE Program.

\end{document}